\documentclass[acmsmall]{acmart} 
\usepackage[inline]{enumitem}
\usepackage{cuted}
\usepackage{graphicx}
\usepackage{subfig}
\usepackage{tikz,lipsum,lmodern}
\usepackage[most]{tcolorbox}
\usepackage{listings}
\usepackage{xcolor}
\usepackage{hyperref}
\usepackage{url}
\usepackage{natbib}
\usepackage{graphicx}
\usepackage{cancel}

\definecolor{codegreen}{rgb}{0,0.6,0}
\definecolor{codegray}{rgb}{0.5,0.5,0.5}
\definecolor{codepurple}{rgb}{0.58,0,0.82}
\definecolor{backcolour}{rgb}{0.95,0.95,0.92}

\usepackage{listings}
\usepackage{color}
\usepackage{fancyvrb}
\definecolor{dkgreen}{rgb}{0,0.6,0}
\definecolor{gray}{rgb}{0.5,0.5,0.5}
\definecolor{mauve}{rgb}{0.58,0,0.82}

\newcommand{\rqone}{How effective is A3Test compared to AthenaTest?}
\newcommand{\rqtwo}{Does A3Test outperform existing pre-trained models?}
\newcommand{\rqthree}{What is the contribution of the assert pre-training and verification components on the performance of A3Test?}
\newcommand{\rqfour}{How efficient is A3Test compared to AthenaTest?}
\newcommand{\smallsection}[1]{ {\bf #1}.\hspace{1mm}}
\newcommand{\ea}{\textit{et al.}}

\usepackage{tikz}
\newcommand*\circled[1]{\tikz[baseline=(char.base)]{
            \node[shape=circle,draw,inner sep=0.5pt] (char) {\small{#1}};}}

\AtBeginDocument{%
  \providecommand\BibTeX{{%
    \normalfont B\kern-0.5em{\scshape i\kern-0.25em b}\kern-0.8em\TeX}}}

\begin{document}

\title{A3Test: Assertion-Augmented Automated Test Case Generation}

\author{Saranya Alagarsamy}
\email{saranya.alagarsamy@monash.edu}

\author{Chakkrit Tantithamthavorn}
\email{chakkrit@monash.edu}

\author{Aldeida Aleti}
\email{aldeida.aleti@monash.edu}
\affiliation{%
  \institution{Monash University}
  \streetaddress{Wellington Road}
  \city{Clayton}
  \state{Victoria}
  \country{Australia}
}


\begin{abstract}
Test case generation is an important activity, yet a time-consuming and laborious task. Recently, AthenaTest---a deep learning approach for generating unit test cases---is proposed. However, AthenaTest can generate less than one-fifth of the test cases correctly, due to a lack of assertion knowledge and test signature verification. In this paper, we propose A3Test, a DL-based test case generation approach that is augmented by assertion knowledge with a mechanism to verify naming consistency and test signatures. A3Test leverages the domain adaptation principles where the goal is to adapt the existing knowledge from an assertion generation task to the test case generation task. We also introduce a verification approach to verify naming consistency and test signatures. Through an evaluation of 5,278 focal methods from the Defects4j dataset, we find that our A3Test (1) achieves 147\% more correct test cases and 15\% more method coverage, with a lower number of generated test cases than AthenaTest; (2) still outperforms the existing pre-trained models for the test case generation task; (3) contributes substantially to performance improvement via our own proposed assertion pre-training and the verification components; (4) is 97.2\% much faster while being more accurate than AthenaTest.

\end{abstract}




\maketitle

\section{Introduction}

Unit testing is a critical component of software development to assure the quality of software systems and improve developers' productivity 
~\cite{cohn2010succeeding}.
However, writing high-quality and effective test case is a difficult and time-consuming task. 
Thus, various automated test case generation approaches are proposed. 
For example, random-based test case generation ~\cite{pacheco2007randoop}, and search-based test case generation~\cite{fraser2011evosuite}.
However, prior studies found that both may achieve good code coverage, but they do not produce human-readable test cases (e.g., generating a test method name as \texttt{test0()}, instead of \texttt{testAddition()}) and the inability to adequately meet the software testing needs of industrial developers ~\cite{almasi2017industrial}, ~\cite{shamshiri2015automated}.
A lack of human-readable test cases could impact the  understanding, debugging, and maintaining activities of the test cases.

Recently, Tufano~\ea~proposed AthenaTest, a Transformer-based model that is learned from developer-written test cases in order to generate correct and readable tests~\cite{tufano2020unit}.
AthenaTest is represented as a translation task where the source is \emph{a focal method} (i.e., the method we would like to test), and the target is \emph{the corresponding test case} originally written by a software developer.
They found that 16.21\% of the generated test cases are correct (i.e., they can correctly test the focal methods and pass the test execution).
Unfortunately, the AthenaTest replication package is not available.

To address this challenge, we first perform a partial replication study (RS) of Tufano~\ea~\cite{tufano2020unit} using Defects4J~\cite{just2014defects4j} as an evaluation dataset.
To do so, we implemented the AthenaTest approach at our own best capability using the hyperparameter settings reported in the original paper.
Unfortunately, AthenaTest fails to generate any correct test cases (0\%), indicating that we are not able to produce the results as reported in the paper.
This has to do with some missing details and settings (e.g., batch size, fine-tuning, epoch).
On the other hand, with our modification, we can successfully implement the AthenaTest approach that achieves results (i.e., 18.08\%) that is similar to  the original paper~\cite{tufano2020unit} (i.e., 16.21\%).
However, the accuracy of the generated test cases is still far from perfect, highlighting the need for further improvement of DL-based test case generation approaches.
In particular, AthenaTest still has the following limitations.
\begin{description}
\item \textbf{Limitation \circled{1}: Lack of assertion knowledge.}
Assertions play an important role in unit testing to assess the expected behaviour. AthenaTest pre-training is limited to natural language datasets and source code. 
This impacts the quality of the generated tests, with 26.71\% of the test cases generated by AthenaTest having incorrect assertions.
For example, \texttt{assertEquals(X, Y, Z)} is an incorrect assertion since \texttt{assertEquals()} should have only two input parameters (i.e., an expected output value and an actual value), \emph{not three}.


\item \textbf{Limitation \circled{2}: Lack of naming  consistency and test signatures verification.}
AthenaTest leverages a general beam search method to generate test cases. 
However, 9.49\% of the test cases generated by AthenaTest are syntactically incorrect. 
For example, AthenaTest may generate an incorrect test method name as \texttt{read()}, where the actual test method name is \texttt{testread()}.
In addition, AthenaTest may generate an incorrect test signature as \texttt{@Test void isLenient()}, where the \texttt{public} keyword is missing from the test method.
\end{description}
Therefore, these limitations may produce  syntactically incorrect, incompatible, and non-readable test cases, which could impact developers' productivity and incur the cost of software testing.













%







In this paper, we propose A3Test
(\underline{A}ssertion \underline{A}ugmented \underline{A}utomated Test Case Generation)
a DL-based test case generation approach that is augmented by assertion knowledge with a mechanism to verify naming consistency and test signatures in order to address the aforementioned limitations of AthenaTest.
First, A3Test leverages the domain adaptation principles where the goal is to adapt the existing knowledge from an assertion generation task to our test case generation task.
To do so, we first build a pre-trained language model of assertions in a self-supervised manner using a PLBART architecture~\cite{ahmad2021unified} with a masked language model.
Therefore, our pre-trained language model is likely to have a stronger foundation knowledge of assertions than AthenaTest.
Then, our pre-trained language model is fine-tuned for the test case generation task where the objective is to learn the relationship between the focal method and the corresponding test case.
For any generated test case, we introduce a verification approach to check the naming consistency (i.e., revising the test method name to be consistent with the focal method name) and the test signatures (i.e., adding missing keywords like \texttt{public}, \texttt{void}, or \texttt{@test} annotations).
Finally, we evaluate our A3Test using a Defects4J dataset ~\cite{just2014defects4j}, which consists of 5K test methods that span across five large-scale open-source software projects (i.e., Lang, Chart, Cli, Csv, Gson) to answer the following four research questions.

\begin{enumerate}[label={\bf(RQ\arabic*)}]

\item {\bf \rqone}\\
\smallsection{Results} 
A3Test achieves 147\% more correct test cases and 15\% more method coverage, with a lower number of generated test cases than AthenaTest.

\item {\bf \rqtwo}\\
\smallsection{Results} 
Not all pre-trained models are effective in the test case generation task.
Nevertheless, A3Test still outperforms the existing pre-trained models (PLBART, CodeGPT, CodeBERT, CodeT5) for the test case generation task.

\item {\bf \rqthree}\\
\smallsection{Results}
Our assertion component contributes to 35.30\%, while our verification component contributes to 23.7\% of the relative improvement when compared to the basic PLBART model. Nevertheless, considering both assertion and verification components perform the best.

\item {\bf \rqfour}\\
\smallsection{Results}
A3Test takes 2.9 hours to generate test cases in one attempt, which is 97.2\% much faster while being more accurate than AthenaTest.

\end{enumerate}


\textbf{Novelty \& Contributions.} 
To the best of our knowledge, we are the first to present:
\begin{enumerate}
    \item An assertion-augmented automated test case generation approach (called A3Test), leveraging the domain adaptation principles, achieving 147\% more correct test cases and 15\% more method coverage than AthenaTest.
    \item Our ablation study shows that each component of our approach contributes 23.70\%-35.30\%, but the combination of both components performs the best.
    \item A replication package of both A3Test and AthenaTest (the baseline approach).
\end{enumerate}




\textbf{Open Science.}
To support the open science community, we publish the studied dataset, scripts (i.e., data processing, model training, and model evaluation), and experimental results in a GitHub repository (\url{http://github.com/awsm-research/a3test}).

\textbf{Paper Organization.} Section 2 describes the problem definition and the limitations of prior work. Section 3 presents our replication study Section 4 presents the A3Test approach. Section 5 presents the experimental setup, while Section 6 presents the results. Section 7 presents an additional discussion. Section 8 discloses the threats to validity. Section 9 draws the conclusions.

\section{Background \& Related Work} \label{sec:background}

\subsection{Unit Testing in a Nutshell} 
Unit Testing is a type of software testing where individual units or components of the software are tested. To perform unit testing, a developer writes a piece of code (unit tests) to verify the code to be tested (unit). Code~\ref{FocalClass} shows an example of a unit test for the \emph{Calculator} class with a single \emph{method} (i.e., Focal Method) and the corresponding test method (i.e., that verifies the method’s behavior with assertions). 

There are software tools and frameworks to support writing and running unit tests, such as Junit~\cite{OMS}, TestNG~\cite{OMSTestng}. JUnit provides methods  such as Mockito ~\cite{Mockito} and assertions to help developers check conditions, outputs, or states in a software program and assess its expected behaviour.
\begin{lstlisting}[language = Java , escapeinside={(*@}{@*)}, , caption= Focal Class Focal Method and Test Case with Assertions, label=FocalClass]
// Focal Class 
public class Calculator{
    // Focal Method
    public double Sum(double first, double second){
        return first + second;
    }
    // Test Method
    @Test 
    public void testSum(){
        double first = 10;
        double second = 20;
        var calculator = new Calculator();
        double result = calculator.Sum(first, second);
        Assert.Equal(30, result); // Assert Statement}
} 
\end{lstlisting}



Our work is connected to a few currently used techniques in the field of automated software testing. Particularly, a class of methodologies—including Evosuite ~\cite{fraser2011evosuite}, Randoop~\cite{pacheco2007randoop} and Agitar~\cite{AgitarTechnologies}—aims to produce test cases. The learning component is the primary distinction between these methods and our strategy.
 
A common approach for automatically generating unit test cases is to do so at random. Randoop~\cite{pacheco2007randoop} checks for errors by generating random sequences of method calls on objects in a Java program and then running these sequences as test cases. Search-based testing is another advanced technique, which uses efficient meta-heuristic search algorithms for test generation. Evosuite~\cite{fraser2011evosuite} is an SBST-based which relies on an evolutionary approach based on a genetic algorithm to generate unit test cases, targeting code coverage. A major weakness and criticism of these approaches is related to the \textit{unsatisfactory code quality} and \textit{understandability} of the generated test cases.

Deep learning-based approaches have been suggested in a few existing studies in the literature for software engineering jobs like code completion~\cite{svyatkovskiy2019pythia} , automatic patch generation ~\cite{tufano2019empirical} ~\cite{chen2019sequencer}, comment generation ~\cite{hu2020deep}, and many others ~\cite{watson2020learning}. 
We incorporate a large amount of uniqueness into this process while also sharing the process of learning from examples with these approaches.
The extensive literature on transfer learning ~\cite{raffel2020exploring}, unsupervised language model pre-training ~\cite{radford2019language}, and denoising pre-training ~\cite{lewis2019bart}~\cite{devlin2018bert} is also relevant to our work.

\subsection{DL-based Test Case Generation}
\label{section:AthenaTest}
Both Random-based and Search-based approaches fall short of the capability to generate readable test cases. Recently, AthenaTest~\cite{tufano2020unit}, a DL-based approach that leverages a sequence-to-sequence BART~\cite{lewis2019bart} transformer model to automatically generate test cases by learning from real-world focal methods and developer-written test cases.
The transformer paradigm, on which AthenaTest is built, seeks to learn the best practices for writing understandable and precise test cases from developer-written test cases. On the other hand, most of the methods currently in use in the literature optimise for code coverage but rely on manually created rules or heuristics to produce test cases.
Transformer-based language models are typically developed in two stages: pre-training and fine-tuning. 

AthenaTest is pre-trained on English~\cite{liu2019roberta} and Java code~\cite{husain2019codesearchnet} and fine-tuned on methods2test~\cite{tufano2022methods2test} dataset. The AthenaTest model takes \emph{focal method as input} and generate \emph{test case as output}. The test case generated by AthenaTest are (i) \textit{realistic:} they resemble developer-written test cases; (ii) \textit{accurate:} they accurately assert the intended behaviour of a focal method, and (iii) \textit{human-readable:} they have readable and understandable code with appropriate variable and method names. However AthenaTest produce low percentage (i.e., 16\%) of correct test cases and requires multiple attempts to generate test cases which emphasis the need for further improved performance of DL-based test case generation.





\begin{table*}

  \caption{(RS) The percentage of correct test cases of AthenaTest from the original paper~\ea~\cite{tufano2020unit}, our replication, and our modification.}
  \label{tab:replicationstudy}
  \resizebox{\linewidth}{!}{%
  \begin{tabular}{lrrrrrrr}
    \toprule
     Model &English Pre-training &GSON &CLI &CSV &CHART &LANG &Total test cases \\
    \midrule
     AthenaTest~\cite{tufano2020unit} & 40 epochs & 2.89\% & 11.07\% & 8.98\% & 11.7\% & 23.35\% & 16.21\%\\
      \hline
     \text{Our Replication}& 40 epochs & 0\% &0\% & 0\%  & 0\% &  0\% &0\% \\
     \text{Our Modification}& 8 epochs & 3.09\% &12.3\% & 8.83\% & 12.5\% & 24.3\% &\textbf{18.08\%} \\
    \bottomrule
  \end{tabular}}
\end{table*}

\begin{table}
    \caption{The hyperparameter settings that are reported by AthenaTest~\cite{tufano2020unit} and the four additional settings (under a horizontal line) that we modified in order to achieve comparable accuracy. }
    \label{tab:hyperparametersetttings}
    \begin{tabular}{lll}
        \toprule
        Hyper-parameters & AthenaTest& AthenaTest' \\
        \midrule
        \text{Encoder, Decoder layers}& 12 & 12 \\
        \text{Learning rate}& 0.0001 & 0.0001 \\
        \text{Optimiser}& Adam & Adam \\
        \text{Code Pre-training Mask}& 20\% & 20\% \\
        \text{DataSet}& Methods2test & Methods2test\\
        \text{Code Pre-Training Epochs}& 10 & 10 \\
        \hline
        Eval Beams& No info & 4 \\
       Batch Size& No info & 32 \\
        English Pre-Training & 40 & 8 \\
        Fine Tuning Epochs & No info & 20 \\
        \bottomrule
    \end{tabular}
\end{table}

\section{A Replication of AthenaTest}
\label{sec:Replication}

In this section, we present our partial replication study (RS) of the AthenaTest approach, proposed by Tufano~\ea~\cite{tufano2020unit}. We employ the Methods2Test~\cite{tufano2022methods2test} dataset to conduct the replication study. Since the code of AthenaTest is not available, we implemented the approach using the parameter settings reported in the original paper.
We use the Defects4J~\cite{just2014defects4j} as an evaluation dataset and evaluate the model using the number of correct test cases, i.e., the test case that passes the execution and invokes the given focal method.

\smallsection{Finding 1} Based on our replication, we find that AthenaTest fails to generate any correct test cases (0\%) (see Table~\ref{tab:replicationstudy}), indicating that we are not able to produce the results as reported in the paper.
We perform a manual analysis of the generated test cases by the AthenaTest approach.
In our replication, we find that the generated test cases are heavily towards natural languages (e.g., \texttt{public class Salim Mehajer ...}), rather than the test cases that should be generated (e.g., \texttt{@Test public void testread()\{...\}}).
We suspect that this should be due to the impact of the pre-training epochs between NL and code corpora.
We observe that, for AthenaTest, the Java pre-training  uses 10 epochs, while the English pre-training uses 40 epochs.
Thus, it could be possible that the pre-training model is learned more towards natural languages than programming languages, generating incorrect test cases towards natural languages.
In addition, we also observe that there are three additional settings (e.g., batch size, fine-tuning, beam search) that are not reported in the paper (see Table~\ref{tab:hyperparametersetttings}).

\smallsection{Finding 2} We can successfully implement the AthenaTest approach that achieves results (i.e., 18.08\%) similar to the original paper~\cite{tufano2020unit} (i.e., 16.21\%).
Since we find that the generated test cases are biased towards natural languages, we reduce the number of epochs for the English pre-training to 8 epochs, which is lower than the Java pre-training to ensure that the model has more influence towards generating Java test cases than generating English text.
Table~\ref{tab:hyperparametersetttings} also reports other parameter settings that we used to achieve comparable results.

\smallsection{Threats to Validity} 
For the replication study, we did not systematically perform parameter tuning, but followed the recommendations from the original study and trialed a few options of possible parameters. The results of our replication can be further improved through parameter tuning. Nevertheless, as shown in Table~\ref{tab:replicationstudy}, our modification performs better than the original AthenaTest by 1.8\%, ensuring that the comparison between our proposed approach and the AthenaTest as a baseline is fair.

\section{A3Test - Assertion Augmented Automated Test Case Generation}



\begin{figure*}[t]
    \centering
    \includegraphics[width=\linewidth]{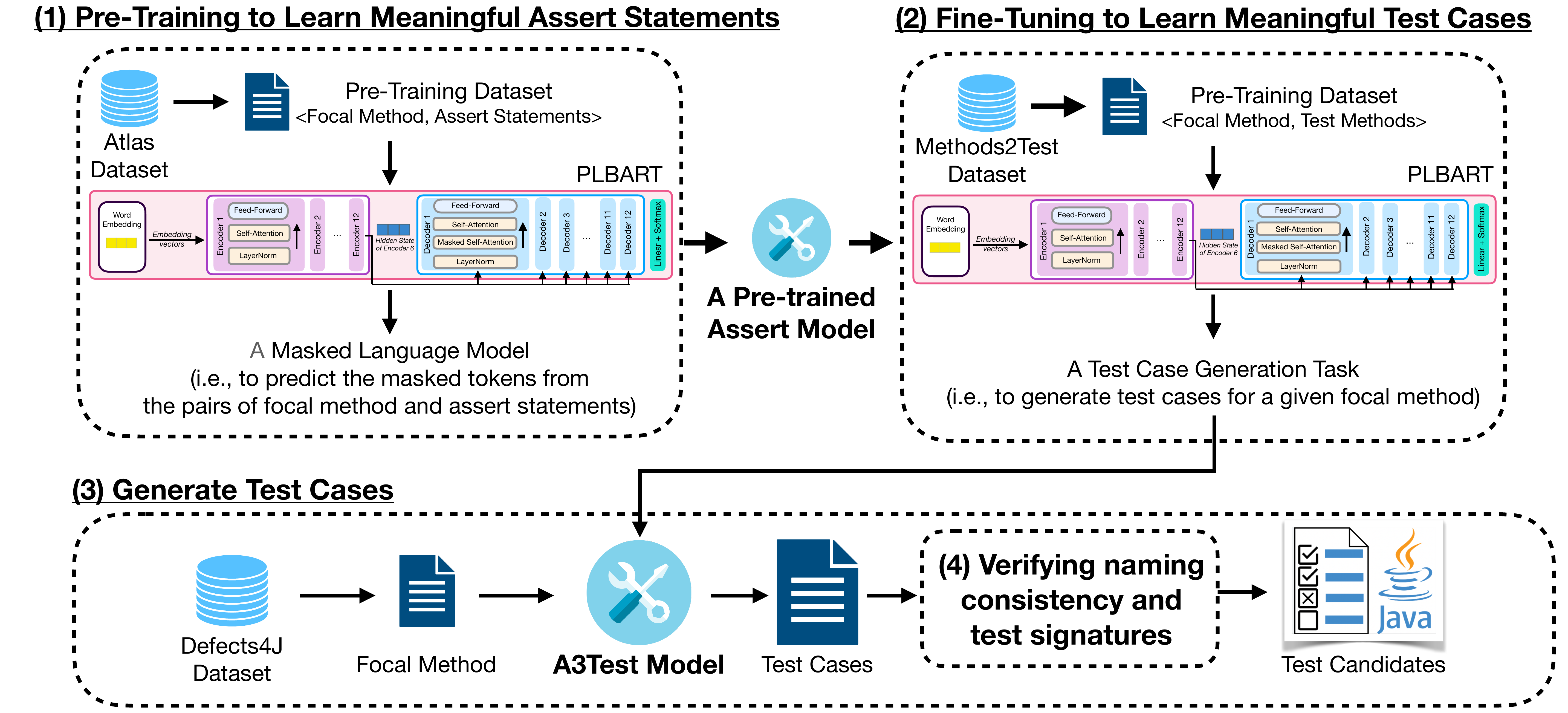}
    \caption{An overview of our A3Test approach, which is a PLBART-based test case generation approach that is augmented by assertion knowledge with a mechanism to verify naming consistency and test signatures.}
    \label{SaranyaPaperModelArch.png}
\end{figure*}

A3Test is a PLBART-based test case generation approach that is augmented by assertion knowledge with a mechanism to verify naming consistency and test signatures. A3Test leverages the domain adaptation principles where the goal is to adapt the existing knowledge from an assertion generation task to the test case generation task.
To do so, we first build a pre-trained language model of assertions in a self-supervision manner using a PLBART architecture~\cite{ahmad2021unified} with a masked language model.
The pre-training learning objective is to predict the masked tokens of a given focal method and the corresponding assert statements.
Therefore, our pre-trained language model is likely to have a stronger foundation knowledge of assertions than AthenaTest.
Then, our pre-trained language model is then fine-tuned with the test case generation task where the objective is to learn the relationship between focal methods and the corresponding test cases.
For any generated test cases, we introduce a verification approach to check the naming consistency (i.e., revising the test method name to be consistent with the focal method name) and the test signatures (i.e., adding missing keywords like \texttt{public}, \texttt{void}, or \texttt{@test} annotations).
We present each step below.

\subsection{Learning Meaningful Assert Statements} 
\label{AssertAugment}

Assertion statements are used to assess the expected behaviour of a unit function.
However, generating test cases is a difficult task since it involves Testing APIs like assertions, which are more than general knowledge of programming languages.
Unfortunately, AthenaTest only builds a pre-trained model based on natural language and source code without considering assertion knowledge.

To address this challenge, we leverage a domain adaptation principle where the goal is to improve the performance of a model on a target domain (i.e., generated test cases) containing insufficient assertion statements by using the knowledge learned by the model from another related domain with adequate labelled data (i.e., generated assertions).
Thus, we define the source domain as an assertion generation (Section \ref{AssertAugment}), where the target domain is a test case generation (Section \ref{FineTuning}). 
The pre-trained assert model is built in a self-supervision manner with a Masked Language Model (MLM).
The pre-training is performed, with the objective to reconstruct the original data from corrupted data and masking 20\% of all tokens. 
MLM (Masked Language Modeling) trains a model to predict a random sample of input tokens that have been replaced by a \texttt{[MASK]} placeholder in a multi-class setting over the entire vocabulary.
The special token \texttt{[MASK]} is replaced for a random sample of the tokens in the input sequence. 
Different from Athena2Test which uses a BART architecture, we use the PLBART architecture~\cite{ahmad2021unified} as a base architecture for building a pre-trained model.
The PLBART architecture leverages a bidirectional attention mechanism to capture context from both the past and future in a sequence, which is different from BART which leverages a standard attention mechanism, allowing our A3Test to efficiently learn the sequence of inputs in parallel.
In addition, PLBART has been pre-trained on a corpus of natural languages and 7 programming languages, while AthenaTest build a BART pre-trained model on a corpus of natural languages and only one programming language.
To build our pre-trained assert model, we use an Atlas dataset ~\cite{watson2020learning}, which is a large corpus of 188,154 pairs of focal methods and assert statements.

\subsection{Learning Meaningful Test Cases}
\label{FineTuning}
Following the domain adaptation principle, we will transfer the pre-trained assert model to fine-tune it in order to learn meaningful test cases.
To do so, we build an A3Test model by fine-tuning the pre-trained assert model with a test case generation task.
The test case generation task is represented as a translation task, with the source being a focal method (i.e., the method we want to test) and the target is the corresponding test case originally written by a software developer.
In particular, we fine-tune our A3Test model using the Methods2Test dataset~\cite{tufano2022methods2test}.
The Methods2Test dataset consists of 780k pairs of a focal method and the corresponding test methods.
We split the dataset into two sets, i.e., training set (80\% - 624,022 pairs) and validation set (10\% - 78,534 pairs).




\subsection{Generating Test Cases}
In the inference phase, we generate a test case for each focal method. 
To do so, we use a beam search~\cite{raychev2014code} as a decoding method.
We leverage beam search to select multiple candidates for an input sequence at each timestep.
Instead of predicting the token with the highest probability at each time step, beam search explores different parts of the search space simultaneously.
The beam search decoding method generates test case candidates while tracking the top-$k$ highest probable candidates (with $k$ being the beam size). 
This allows beam search to select the best candidates with the highest probability using a best-first search strategy for generating test cases, allowing our A3Test to efficiently generate
a single best test case instead of generating 30 candidates like AthenaTest.







\subsection{Verifying Naming Consistency and Test Signatures}
\label{verification}

It is possible that the generated test cases could be syntactically incorrect (e.g., generating incomplete parenthesis, incorrect test method names or invalid test signature keywords), which impacts the performance of our A3Test.
To address this challenge, we introduce an automated verification approach (which was not previously done by AthenaTest) in order to check and correct the naming consistency and invalid test signatures.
Our verification approach consists of three parts: (1) verify the incomplete parenthesis, (2) verify the naming consistency and (3) verify the test signatures.

\circled{1} \textbf{Verifying the incomplete parenthesis.} 
It is possible that the generated test cases may have an incomplete parenthesis (e.g., (), \{\}).
Thus, we develop an approach to detect and correct missing parenthesis using a push-pop algorithm.
Such an approach allows us to detect and correct the missing parenthesis.

\circled{2} \textbf{Verifying the naming consistency.} 
In general, test method names must adhere to the naming convention of the JUnit framework (e.g., a test method name must start with \texttt{test}).
Thus, a test case may be incorrect due to an incorrect method name (aka. naming inconsistency).
For example, a test method name as \texttt{read()} is considered incorrect, since it does not start with \texttt{test}.
Therefore, the testing framework will not recognize it as a test case and will not execute it.
We develop an approach to detect and correct naming consistency using a string processing approach.
We check the prefix of the method names (i.e., check the first four letters whether it contains \texttt{test} or not).
Our approach will add the prefix \texttt{test} to the test method name if the prefix is missing.
Thus, based on a given example, the incorrect test method name (\texttt{read()}) will be automatically revised to \texttt{testread()}.

\circled{3} \textbf{Verify the test signatures.}
Test cases will not be executed if they do not match the test method signatures of the JUnit framework.
In general, Java methods can be private, public, protected or package-private. 
However, test cases are encouraged to be a public method only in order to be executed.
Unfortunately, DL-based test case generation approaches are not specifically trained to generate public methods only, but could be others too.
In addition, there was a lack of a mechanism to verify the test signatures in order to be successfully executed.
Thus, we introduce an approach to verify the test signatures based on a string processing approach.
For a given test method, we first check the sequence and the existence of the first four tokens (i.e., \texttt{@Test public void test[MethodName]\{...\}}) if the following four specific keywords exist or not, i.e., \texttt{@Test}, \texttt{public}, \texttt{void}, and \texttt{test}.
Any keywords that are missing from the sequence of the first four tokens will be automatically added to the generated test methods to ensure that the generated test methods by our A3Test will be successfully executed.

\section{EXPERIMENTAL SETUP}

In this section, we present the detail of our experimental setup, including datasets, model implementation, model training, and hyperparameter settings.

\textbf{Datasets.} Similar to AthenaTest~\cite{tufano2020unit}, we use Defects4J~\cite{just2014defects4j} as a benchmark evaluation dataset.
Defects4J provides a collection of real-world open-source Java projects that can be used to evaluate and compare various software testing techniques. 
Each project is a collection of Java programs, including java class files and the focal method.
In particular, we chose the same five Defect4j projects as AthenaTest for test case generation, namely, Apache Commons Lang ~\cite{Apache}, JFreeChart~\cite{JFreeChart}, Apache Common CLI ~\cite{ApacheCommonsCLI}, Apache Common CSV~\cite{ApacheCommonsCSv}, Google Gson ~\cite{googleGson}.
Subsequently, we parse the focal classes and extract the list of every public method for each project. 
In summary, we have a total of 5,278 focal methods, including 2,712 focal methods from Lang, 1,328 focal methods from Chart, 645 focal methods from Cli, 373 focal methods from Csv, and 220 focal methods from Gson. 
Each one of these public methods represents a focal method for which we aim to generate test cases. 

%



\textbf{Model Implementation.} A3Test is built on top of two deep-learning Python libraries, Transformers~\cite{wolf2019huggingface} and PyTorch~\cite{collobert2011torch7}. 
The Transformers library provides API access to transformer-based model architectures, while the PyTorch library aids in computation during training.

\textbf{Model Training.} The PLBART tokenizer and model pre-trained by Ahmad~\ea~\cite{ahmad2021unified} are obtained from the Transformers library. 
To generate test cases, we use the methods2test dataset to fine-tune our pre-trained assert model. 
For the AthenaTest approach, we use all the best hyperparameters described in Tufano~\ea~\cite{tufano2020unit}.
The experiment is run on one NVIDIA GeForce RTX 3090 GPU with 24 GB memory, an Intel(R) Core(TM) i9-9980XE CPU@3.00GHz with 36 core processors, and 64G RAM. 

\textbf{Hyper-Parameter Settings for Fine-Tuning.} For the model architecture of our A3Test approach, we use the default setting of the PLBART base model (i.e., 12 Transformer Encoder/Decoder blocks, and 12 attention heads). 
During fine-tuning, the learning rate is set to $1e^{-5}$ with a linear schedule. 
We use AdamW optimizer~\cite{loshchilov2017decoupled} which is widely adopted to fine-tune our A3Test models to update the model and minimise the loss function.

\section{Experimental Results}
\subsection*{RQ1: \rqone}

\begin{table}[t]
    \caption{(RQ1) The percentage for the correct test cases and the focal method coverage of A3Test and AthenaTest.}
      \renewcommand{\arraystretch}{1.2}
    \label{tab:CorrectTest}
    \begin{tabular}{l | r | r | r | r}
        \hline
 & \multicolumn{2}{| l |}{\#Correct Test Cases} & \multicolumn{2}{| l}{Focal Method Coverage} \\ \hline
    Projects    & A3Test           & AthenaTest         & A3Test            & AthenaTest            \\ \hline
GSON    & 14.09\%          & 2.89\%             & 40.90\%            & 9.54\%                \\ 
CLI     & 25.19\%          & 11.07\%            & 37.20\%            & 29.46\%               \\ 
CSV     & 25.73\%          & 8.98\%             & 37.80\%            & 34.31\%               \\ 
CHART   & 31.30\%           & 11.70\%             & 34.40\%            & 32.00\%               \\ 
LANG    & 49.50\%           & 23.35\%            & 58.30\%            & 56.97\%               \\ \hline
Total   & \textbf{40.05\%}          & 16.21\%            & \textbf{46.80\%}            & 43.75\%     \\ \hline    
\end{tabular}
\end{table}


\textbf{Motivation.} Tufano~\ea~\cite{tufano2020unit} proposed AthenaTest for test case generation.
However, there were some limitations that remain unexplored (e.g., lack of assertion knowledge and lack of the naming consistency and test signatures verification).
To address this challenge, we propose A3Test, a DL-based test case generation approach that is augmented by assertion knowledge with a mechanism to verify naming consistency and test signatures in order to address the limitations of AthenaTest.
Thus, we formulate this RQ to investigate the effectiveness of A3Test when compared to AthenaTest.

\textbf{Approach.} To answer this RQ, we evaluate the effectiveness of our A3Test approach and compare it with AthenaTest using the following two evaluation measures, similar to Tufano~\ea~\cite{tufano2020unit}.
First, we use \textbf{the number of correct test cases} to measure the number of passing test cases that cover the given focal method.
Second, we also use \textbf{a focal method coverage} to measure the number of focal methods that are covered by at least one of the generated test cases.
To do so, we execute the test cases through the JUnit~\cite{OMS} framework in order to obtain a test coverage analysis report. 
In the report, we will be able to identify (1) the passing test cases and (2) the covered focal methods.
Knowing the number of correct test cases and the focal method coverage allows developers to make a better data-informed decision about whether the approach is effective or not.
Finally, we compute the relative percentage improvement of each measure ($m$) between our approach and the baseline as follows:

\begin{equation}
    \frac{m_\textrm{A3Test}}{m_\textrm{A3Test}-m_\textrm{AthenaTest}} \times 100\%.
\end{equation}

\textbf{Results.} \textbf{A3Test generates 147\% more correct test cases and 15\% more method coverage with a lower number of generated test cases than AthenaTest.}
Table~\ref{tab:CorrectTest} presents the number of correct test cases and the focal method coverage of A3Test and AthenaTest.
For the number of correct test cases, our A3Test achieves 40.05\%, meaning that 40.05\% of the generated test cases are correct.
On the other hand, AthenaTest achieves as low as 16.21\%, meaning that only 16.21\% of the generated test cases are correct.
This is also consistent with each individual Defects4J project, as we find that our A3Test approach consistently performs 112\%-387\% better than AthenaTest in terms of the number of correct test cases.
For the focal method coverage, our A3Test achieves 46.80\%, meaning that the generated test cases by our A3Test can cover 46.80\% of the focal methods.
On the other hand, AthenaTest achieves 43.75\%, meaning that the generated test cases by our A3Test can cover 43.75\% of the focal methods
This is also consistent with each individual Defects4J project, as we find that our A3Test approach consistently
2\%-411\% performs better than AthenaTest in terms of the focal method coverage.

\textbf{A3Test achieves a higher number of correct test cases and higher method coverage, with a lower number of generated test cases than AthenaTest.}
Ideally, a highly effective test case generation approach should generate a minimal set of test cases that covers a maximal set of focal methods. A3Test and AthenaTest have different internal pre-training mechanisms, producing a different number of generated test cases. AthenaTest leverages an explicit pre-training strategy (i.e., building a BART pre-training by themselves on the English/Code corpus). AthenaTest is designed to generate test cases with 30 attempts for each focal method, generating a total of 158,400 test cases (i.e., 5,278 focal methods $\times$ 30 attempts). Only 16.21\% ($\frac{25,680}{158,400}$) of the generated test cases are correct, which covers up to 43.75\% of the focal methods.
Different from AthenaTest, A3Test requires a single attempt to generate a test case.
That means A3Test generates a total of 5,278 test cases (i.e., 5,278 focal methods $\times$ 1 attempt) where 40.05\% of the generated test cases ($\frac{2,114}{5,278}$) are correct, which covers up to 46.80\% of the focal methods.
Despite the lower number of generated test cases, A3Test achieves a higher number of correct test cases and higher method coverage, highlighting the significant upper hand of A3Test.

\begin{tcolorbox}[colback=gray!5!white,colframe=gray!75!black]
Our A3Test achieves 147\% higher number of correct test cases and 15\% higher focal method coverage, with a lower number of generated test cases than AthenaTest.
\end{tcolorbox}

\subsection*{RQ2: \rqtwo}

\textbf{Motivation.} The pre-training component plays an important role in test case understanding and generation.
However, different test case generation approaches have different pre-training strategies.
For example, AthenaTest leverages a BART architecture to build its own pre-trained models via an explicit pre-training strategy. 
On the other hand, A3Test leverages a PLBART architecture as a base model via an implicit pre-training strategy. 
Also, there exist other pre-trained language models of code (e.g., CodeT5~\cite{wang2021codet5}, CodeBERT~\cite{feng2020codebert}, CodeGPT~\cite{lu2021codexglue}).
These pre-trained models have been successfully used in various software engineering tasks, e.g., code completion, code summarization, code generation, and code transformation, \emph{but not for test case generation}.
Thus, it remains unclear which pre-trained models are the best for test case generation and whether our A3Test outperforms the standard pre-trained models or not.
Therefore, we formulate this RQ to investigate the performance of various pre-trained models when compared to our A3Test.



\textbf{Approach.} To answer this RQ, we select the four existing pre-trained models of code, namely, CodeT5~\cite{wang2021codet5}, CodeBERT~\cite{feng2020codebert}, CodeGPT~\cite{lu2021codexglue}, and PLBART~\cite{ahmad2021unified} as a base model for the test case generation task, without including the other components (Assert+Verification).
Then, we compare the performance of these models with our A3Test and AthenaTest.
Finally, we evaluate the performance of the models using the number of correct test cases.



\textbf{Results.} 
\textbf{Not all pre-trained models are effective in the test case generation task.} 
Table~\ref{tab:comparision} presents the performance of A3Test and compares it with different pre-trained language models.
We find that the performance of the existing pre-trained models varies greatly from 0\% (CodeGPT, CodeBERT) to 21.5\% (PLBART) for the test case generation task.
This finding indicates that different pre-trained models are task-specific. 
While pre-trained models have been successfully used in various software engineering tasks, e.g., code completion, code summarization, code generation, and code transformation, they do not imply that they will perform best in the test case generation task.
This finding highlights the importance of investigating different pre-trained models prior to adopting them for the downstream task.

Nevertheless, \textbf{our A3Test still outperforms the existing pre-trained models (PLBART, CodeGPT, CodeBERT) for the test case generation task.} 
When comparing between A3Test (PLBART + Assert + Verification) and PLBART alone, we find that our A3Test still performs better than PLBART alone (i.e., improving from 21.50\% to 40.05\%), confirming that using PLBART alone is not effective enough for the test case generation task.
In addition, the PLBART alone still performs better than AthenaTest (the state-of-the-art approach), improving from 16.21\% to 21.50\%.
While both AthenaTest and PLBART are based on the same BART architecture, their pre-training strategies are different.
This finding confirms that PLBART which leverages the implicit pre-training strategy performs better than the explicit pre-training strategy used by AthenaTest.
This highlights the benefit of our A3Test that leverages PLBART instead of using a basic BART architecture.

\begin{table*}[t]
    \caption{ (RQ2) The performance of A3Test and AthenaTest when compared to other pre-trained language models (measured by the percentage for the correct test cases).}
  \label{tab:comparision}
    \renewcommand{\arraystretch}{1.2}
  \begin{tabular}{l | r | rrrrr}
    \toprule
     \textbf{Model} & \textbf{\%Correct} & GSON & CLI & CSV & CHART & LANG  \\
    \midrule      
    A3Test & 40.05\% & 14.09\% & 25.19\% & 25.73\% & 31.30\% & 49.50\% \\
    AthenaTest & 16.21\% & 2.89\% & 11.07\% & 8.98\% & 11.70\% & 23.35\% \\
    \hline
     PLBART & 21.50\% & 5.04\% & 13.70\% & 10.10\%   & 9.26\%   & 22.40\%  \\
    CodeT5 & 13.90\%  &  3.63\%   & 11.00\%  & 8.57\% &  7.60\% & 19.28\% \\
    CodeBERT& 0\% &0\%  & 0\% &0\%  &0\%  &0\%\\
    CodeGPT&  0\% &0\%  & 0\% &0\%  &0\%  &0\%\\
      
    \bottomrule
  \end{tabular}
\end{table*}

\begin{tcolorbox}
[colback=gray!5!white,colframe=gray!75!black]
Not all pre-trained models are effective in the test case generation task.
Nevertheless, our A3Test still outperforms the existing pre-trained models (PLBART, CodeGPT, CodeBERT) for the test case generation task.
\end{tcolorbox}

\subsection*{RQ3: \rqthree}

\textbf{Motivation.} Our A3Test approach consists of three key components, i.e., PLBART + Assert Pre-Training + Verification.
However, little is known about which components contribute the most to the performance of our A3Test.
Thus, we formulate this RQ to conduct an ablation study to investigate the performance of the components of our A3Test approach.

\textbf{Approach.} We conduct an ablation study to investigate the performance of the components of our A3Test approach.
We extend our experiment to systematically evaluate the following four variants of A3Test by removing the Assertion Pre-Training and Verification components as follows:

\begin{itemize}
\item \texttt{PLBART}: The PLBART architecture without the assertion pre-training and verification components.
\item \texttt{PLBART+Verification}: The PLBART architecture with the verification component, but without the assertion pre-training component.
\item \texttt{PLBART+Assert}: The PLBART architecture with the assertion pre-training component, but without the verification component.
\item \texttt{PLBART+Assert+Verification}: Our own A3Test.
\end{itemize}


 \begin{table*}[ht]
    \renewcommand{\arraystretch}{1.2}
    \caption{(RQ3) The percentage of correct test cases generated by each variant of A3Test.}
    \label{tab:A3Test}
    \begin{tabular}{l | r | rrrrr}
    \toprule
    \textbf{Model} & \textbf{\%Correct} & GSON & CLI &CSV &CHART &LANG  \\
    \midrule       
    PLBART &21.50\% &5.04\% & 13.70\% &10.10\% &9.26\%  &22.40\%  \\
    PLBART+Assertion &29.10\% & 8.60\%&25.19\%  & 19.03\% &21.30\%  &36.70\%  \\     
    PLBART+Verification &26.60\% & 11.40\% &12.30\% & 24.04\% &19.60\%  &45.80\%  \\
    A3Test & \textbf{40.05\%} & 14.09\% &35.19\% & 25.73\%  &31.30\% &49.50\% \\ 
    \bottomrule
    \end{tabular}
\end{table*}

\textbf{Results.}
\textbf{The assertion component of A3Test contributes 35.30\%, while the verification component contributes 23.7\% of the relative improvement when compared to the basic PLBART model.}
When comparing PLBART and PLBART + Assertion, we find that the performance is improved from 21.50\% to 29.10\%, contributing to 35.30\% of the relative improvement.
On the other hand, when comparing PLBART and PLBART + Verification, we find that the performance is improved from 21.50\% to 26.60\%, contributing to 23.7\% of the relative improvement.
These findings highlight that each of our own proposed components substantially contributes to the performance improvement of the A3Test approach.

Nevertheless, \textbf{our A3Test approach that considers both assertion and verification components still perform the best.}
Although we find that each component can contribute to performance improvement to some extent, when considering both assertion and verification components, the performance is improved from 21.50\% to 40.05\%, accounting for 86.2\% ($\frac{40.05}{40.05-21.50}$) of the improvement, highlighting the importance of our own proposed assertion and verification component for test case generation.





\begin{tcolorbox}[colback=gray!5!white,colframe=gray!75!black]
The assertion component of A3Test contributes to 35.30\%, while our verification component contributes to 23.7\% of the relative improvement when compared to the basic PLBART model. Nevertheless, considering both assertion and verification components perform the best.
\end{tcolorbox}

\subsection*{RQ4: \rqfour}

\smallsection{Motivation}
The efficiency of the test case generation approaches is an important perspective to consider about the adoption of research-driven approaches in practice.
Thus, we formulate this RQ to investigate what is the efficiency of our A3Test approach when compared to AthenaTest.


\smallsection{Approach}
Different environments may produce different execution times, which may impact the efficiency of the test case generation approaches.
To ensure a fair comparison, we decide to run the AthenaTest (the modification version in Section~\ref{sec:Replication}) in our environment, which is the same as we run our A3Test approach.
We also run each approach individually to ensure that the time measurement is accurate.
For each focal method, we measure the computational time that each approach takes to generate test cases.
Finally, we report the average time to generate each test case and the total amount of time to generate test cases for all of the 5,278 focal methods.


\smallsection{Results} \textbf{A3Test takes 2.9 hours to generate test cases in one attempt, which is 97.2\% faster while being more accurate than AthenaTest.}
Among the total 5,278 test cases (see Table~\ref{tab:time}), A3Test takes 2.9 hours and requires only 1 attempt, while AthenaTest takes 105 hours and requires 30 attempts, which results in a 97.2\% efficiency improvement for test case generation.
One attempt of AthenaTest takes 3.5 hours. The average time to generate each test case by A3Test is 1.98 seconds, while AthenaTest requires on average 2.34 seconds.
These findings confirm that our A3Test is considerably more efficient and generates more correct test cases than AthenaTest.

 \begin{table}[t]

  \caption{(RQ4) The computational time to generate one test case (on average), to generate test cases in one attempt and in 30 attempts for all of the 5 studied Defects4J projects.}
  \label{tab:time}
    \renewcommand{\arraystretch}{1.2}
  \begin{tabular}{l | rrr}
    \toprule
     \textbf{Model} & \textbf{1 Test Case (Avg)} & \textbf{1 Attempt} & \textbf{30 Attempts} \\
    \midrule            
     \text{A3Test}&1.98 seconds & 2.9 hours & - \\ 
     \text{AthenaTest}& 2.34 seconds & 3.5 hours &105 hours\\
       
    \bottomrule
  \end{tabular}
\end{table}

\begin{tcolorbox}[colback=gray!5!white,colframe=gray!75!black]
Our A3Test takes 2.9 hours to generate test cases in one attempt, which is 97.2\% much faster while being more accurate than AthenaTest.
\end{tcolorbox}

\section{DISCUSSION \& FUTURE WORK}


Our early analysis reveals favourable findings in a few areas. Our method can use several testing APIs such as AthenaTest and build syntactically accurate test cases that adhere to the test case standards. The generated test cases appear to be (i) accurate — correctly asserting the expected behaviour of a focal method; and (ii) human-readable — readable and understandable code with appropriate variable and method names. However, more analysis needs to be performed.
 
By moving away from coverage-driven approaches and towards machine learning models that seek to comprehend code, we think our work represents a preferable approach to the new category of automated test case generation tools. These learning techniques could result in test cases that are more naturally produced, and better fit the existing code base.
\subsection{Line Coverage Analysis}

Our A3Test results reveal that our approach was able to generate correct test cases (40.05\%) with the best line coverage than AthenaTest.
 We analysed the line coverage for \textit{Class NumberUtils of Lang-1-f}.
 We use this Class file as a motivating example of our A3Test vs AthenaTest to understand the quality of generated tests.

Table ~\ref{tab:coverage} shows the results of our line coverage analysis comparing A3Test and AthenaTest.
The table shows the absolute (and percentage) of line coverage for each of the 18 unique public methods, with the best coverage value highlighted in bold and the same results marked with an underline. Compared to AthenaTest, our A3Test approach is able to generate correct test cases with the best line coverage for most of the focal methods. 

\begin{table}[ht]
  \caption{Test Coverage Analysis – Test cases generated by AthenaTest and A3Test are executed and their coverage is analyzed in terms of lines covered.}
  \label{tab:coverage}
    \renewcommand{\arraystretch}{1.2}
  \begin{tabular}{lrr}
    \toprule
     Focal Method &AthenaTest &A3Test\\
    \midrule
     \text toInt(String, int) &23 (6.1\%) &\textbf{24 (6.4\%)}\\
      \text toLong(String, long) &20 (5.3\%) &\textbf{21 (5.6\%)}\\
        \text toFloat(String, float) &22 (5.9\%) &21 (5.6\%)\\
    \text toDouble(String, double) &20 (5.3\%) &\textbf{21 (5.6\%)}\\
    \text toByte(String, byte) &23 (6.1\%) &\underline{23 (6.1\%)}\\
    \text toShort(String, short) &22 (5.9\%) &\textbf{23 (6.1\%)}\\
    \text createFloat(String) &20 (5.3\%) &\textbf{21 (5.6\%)}  \\
    \text createDouble(String)  &21 (5.6\%) &\underline{21 (5.6\%)}\\
    \text createInteger(String) &21 (5.5\%) &-\\
    \text createLong(String)  &21 (5.5\%) &\textbf{23 (6.1\%)}\\
    \text createBigInteger(String)  &20 (5.3\%) &\textbf{28 (7.5\%)}\\
    \text createBigDecimal(String)  &22 (5.9\%) &\underline{22 (5.9\%)}\\
    \text min(long[]) &22 (5.9\%) &\underline{22 (5.9\%)}\\
    \text min(int, int, int) &22 (5.9\%) &\textbf{25 (6.7\%)}\\
    \text max(float[])  &22 (5.8\%) &\textbf{23 (6.1\%)}\\
    \text max(byte, byte, byte) &22 (5.9\%) &\underline{22 (5.9\%)}\\
    \text isDigits(String)  &23 (6.1\%) &\underline{23 (6.1\%)}\\
    \text isNumber(String)  &51 (13.6\%) &33 (8.8\%)\\
    \bottomrule
  \end{tabular}
\end{table}
  



\subsection{Generation calls to private fields}

We further analysed the falling test cases. Our model was able to predict test cases correctly around the focal context information such as class names, constructors, other method signatures, and fields associated with the focal method. 

\begin{figure}[h]
  \centering
  \includegraphics[width=\linewidth]{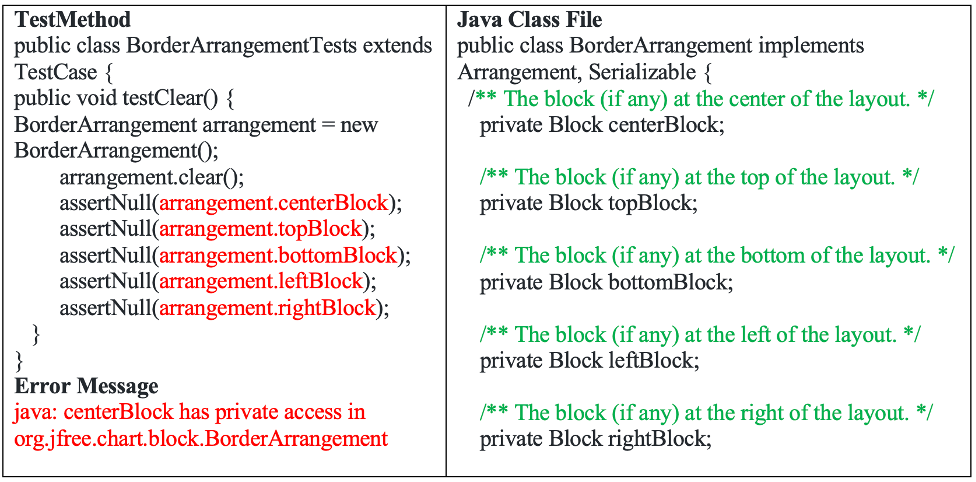}
  \caption{Examples of Test Cases with private field}
  \label{test casePrivateFeild}
\end{figure}

Figure~\ref{test casePrivateFeild} shows an example of generated test cases for the class BorderArrangement of Chart-13-f, our model generated test case using the private fields. The test case execution failed because of access restriction of the fields. This implies that our approach is capable of generating correct test cases, with assertions based on focal context details. These failed test cases can be considered as an example to verify if the access modifiers work as expected. The failing test is considered a “positive” however, a “positive” does not necessarily indicate that the oracle caught the bug. A failing test can indicate that:
\textbf{True Positive} - The test has a correct oracle and fails due to the class implementation. These test cases are not taken into account in the current research but serve as inspiration for future work, with the goal of improving our model.

\subsection{Deprecated Assert Statements.} 
We observed that in A3Test 6\% to 8\% of the cases generate test cases using \texttt{Assert.assertThat()} method, which is a deprecated test method in JUnit version 5.
The current version of JUnit Jupiter’s Assertions class does not provide an \texttt{Assert.assertThat()} method like the one found in JUnit 4’s \texttt{org.junit}. 
We believe that our model was fine-tuned on the methods2test dataset, which includes test cases that utilize the \texttt{Assert.assertThat()} method.
We will perform a further investigation in the future to mitigate this challenge.



\section{Threats to Validity}
As for any empirical study, there are various threats to the validity
of our results and conclusions.

\textbf{Threats to the internal validity} related to the degree to which our
study minimizes systematic error. Our AthenaTest replication consists of various
hyperparameter settings (i.e., number of hidden layers, number
of attention heads, and learning rate). Prior studies raise concerns
that different hyperparameter settings may have an impact on the
evaluation results.
However, finding an optimal hyperparameter setting can be very expensive
given a large search space of the Transformer architecture.
Instead, the goal of our work is not to find the best hyperparameter
setting, but to fairly compare the accuracy of our approach with
the existing baseline approaches. 
Thus, the accuracy reported in
the paper serves as a lower bound of our approach, which can be even further improved through hyperparameter optimization.
To mitigate this threat, we report the hyperparameter settings of our
replication package to aid future replication studies.

\textbf{Threats to external validity} concern the generalization of our findings. Experiments are based on five projects from Defects4J~\cite{just2014defects4j}. This is in line with the prior study on AthenaTest. To circumvent the threat to external validity, projects were selected with diversity in mind. The five projects represent different domains of inputs (string, int, etc.) and sizes and complexity of classes. Further experiments with other projects would help with the generalisability of the results.

\textbf{Threats to construct validity} concern the relation between experimentation and theory. We have compared the performance of the testing techniques based on \textit{method coverage}, which is a widely used performance metric in the literature. However, in the future, it is worth reporting the performance based on other metrics, e.g. mutation score, fault detection etc. We are very interested in such research as future work. 

\section{CONCLUSION}
In this paper, we propose A3Test, a pre-trained Transformer-based approach that is augmented by assertion knowledge with a
mechanism to verify naming consistency and test signatures.
We discovered that our A3Test method outperformed AthenaTest in several areas, after analyzing 5,278 focal methods from the Defects4j dataset. 
Specifically, A3Test generated 147\% more correct test cases and achieved 15\% more method coverage, while using fewer test cases than AthenaTest. 
A3Test also surpassed existing pre-trained models such as PLBART, CodeGPT, CodeBERT, and CodeT5 for test case generation.
Our proposed assertion pre-training and verification components played a significant role in performance improvement. Moreover, A3Test was much faster than AthenaTest, with a speed improvement of 97.2\% while maintaining higher accuracy.
Our results confirm that A3Test is more accurate in generating correct test cases. 
We therefore, anticipate that our approach could assist developers in producing effective and efficient test code.

\section*{Acknowledgement}
Chakkrit Tantithamthavorn was partly supported by the Australian Research Council's Discovery Early Career Researcher Award (DECRA) (DE200100941).

\bibliographystyle{ACM-Reference-Format}
\bibliography{sample-base}
\end{document}